\documentclass[]{aa}

\usepackage[varg]{txfonts}
\usepackage{amssymb,amsmath,amsfonts,mathrsfs}
\usepackage{xcolor}
\usepackage[colorlinks=true]{hyperref}
\hypersetup{citecolor=blue,linkcolor=magenta}
\usepackage{multirow,array}
\DeclareMathAlphabet{\pazocal}{OMS}{zplm}{m}{n}
\newcommand{\gtwo}{{}_{2}g_{2}}
\newcommand{\torf}{{}_{4}t_{0}}

\newcommand{\tor}{{}_{2}t_{0}}

\begin{document} 

\titlerunning{QPOs in precursor flares via seismic aftershocks}
   \title{Quasi-periodic oscillations in precursor flares via seismic aftershocks from resonant shattering}

   \author{A. G. Suvorov\inst{1,2}\fnmsep\thanks{arthur.suvorov@tat.uni-tuebingen.de}
          \and
          H. J. Kuan\inst{1,3}
          \and
          K. D. Kokkotas\inst{1}
          }

   \institute{Theoretical Astrophysics, IAAT, University of T{\"u}bingen, T{\"u}bingen, D-72076, Germany
         \and
             Manly Astrophysics, 15/41-42 East Esplanade, Manly, NSW 2095, Australia
             \and
             Department of Physics, National Tsing Hua University, Hsinchu 300, Taiwan
             }

   \date{Received 22/05/2022; accepted 30/06/2022}

% \abstract{}{}{}{}{} 
% 5 {} token are mandatory
 
  \abstract
   {GRB 211211A was accompanied by a kilonova, implying a merger origin for the event. A precursor flare, modulated by  quasi-periodic oscillations at $\approx 22\, \mbox{Hz}$, was also seen $\approx 1\, \mbox{s}$ prior to the main emission.}
   {It is suggested here that the precursor resulted from the resonant shattering of one star's crust prior to coalescence.
   Seismic aftershocks and low-frequency torsional modes may emanate from the rupture site in this case, explaining the underlying oscillations. This interpretation is directly analogous to proposals for giant flares from soft gamma repeaters, where oscillations at similar frequencies have been observed, involving starquakes followed by crustal vibrations.}
   {Torsional mode properties were computed for sequences of slowly rotating, magnetised neutron stars in general relativity.}
   {The $\approx 22\, \mbox{Hz}$ modulations in the precursor matches that of nodeless, $\ell =2$ torsional modes for a wide variety of stellar parameters. {Global or discrete Alfv{\'e}n modes are also viable candidates.} }
   {}

   \keywords{gamma-ray bursts: 211211A -- stars: magnetars -- stars: oscillations
               }

   \maketitle

\section{Introduction}
\label{sec:intro}

Gamma-ray bursts (GRBs) are observationally classified as being either long or short depending on their duration, $T_{90}$. Long GRBs are typically associated with supernovae, and in most cases are well-explained by a `collapsar' model where a relativistic jet, which formed at the heart of a core collapse, comes to drill through the stellar envelope, eventually producing the observed gamma rays \cite[e.g.][]{bloom02,omer12}. Short GRBs instead arise from binary merger events involving at least one neutron star, and they are often associated with kilonovae \cite[e.g.][]{cio18}. GRB 211211A, however, challenges this simple dichotomy \cite[see also][for the case of GRB 060614]{zhang07}, as the event was not accompanied by a supernova but rather a kilonova, despite its long duration \cite[$T_{90} \sim 35\, \mbox{s}$;][]{gomp22,rast22,yang22}. This suggests that GRB 211211A originated from a merger.

GRB 211211A was preceded by a non-thermal, soft gamma-ray flash (`precursor flare') $1.08 \pm 0.20\, \mbox{s}$ prior to the main event \citep{xiao22}. This precursor flare, which is spectrally similar to those observed in short GRBs \citep{wang20}, may have launched prior to coalescence at an orbital separation that depends on a sub-second jet-formation timescale, $\tau_{\rm jet}$ \citep{zhang19}. For the first time, however, quasi-periodic oscillations (QPOs) at $22^{+3}_{-2}\, \mbox{Hz}$ were detected within the precursor flare at the $>5\sigma$ level \citep{xiao22}. Although less significant, a second Fourier peak is also visible at $51 \pm 2\, \mbox{Hz}$ in the Fermi data \cite[see extended data Figure 7 in][]{xiao22}.

Tidal resonances may be responsible for precursor flares from binary mergers \citep{tsang12,tsang13}. The basic mechanism involves the excitation, via orbital resonances, of stellar quasi-normal modes (QNMs) within one of the inspiralling constituents. If polar-parity QNM amplitudes reach a sufficiently high peak during the resonance window, the crust may strain enough so that it fractures, releasing copious gamma rays. Depending on the orbital frequency at the time of the precursor, different modes may be favoured, such as $f$ \citep{suvk20}, $i$ \citep{tsang21}, or $g$ modes \citep{kuan21b}.

Crustquakes are also popular models for giant flares in soft gamma repeaters \cite[SGRs;][]{td95}. As first argued by \cite{duncan98} \cite[see also][]{blaes89}, global seismic oscillations are likely to accompany a fracture; torsional modes, which are restored by the weak Coulomb forces of crustal ions, are especially easy to excite in such a scenario. Torsional mode properties have been calculated by many authors at various levels of sophistication. General-relativistic \citep{schu83}, rotational \citep{lee96,vav07}, and especially magnetic \citep{mess01,sot07,col12,link16} effects, which are most relevant for magnetars, have been studied. The torsional eigenfunctions primarily depend on the shear modulus of the crust, $\mu$ \cite[e.g.][]{tews17}. Studies agree, however, that the nodeless, $\ell =2$ modes ($\tor$) for slowly rotating and modestly magnetised stars have frequencies in the range of $20 \lesssim \tor/\text{Hz} \lesssim 30$; $\ell =4$ mode frequencies instead have $40 \lesssim \torf/\text{Hz} \lesssim 60$. If quakes are responsible for precursor flares, it is therefore natural that damped oscillations (i.e. QPOs) at these frequencies manifest within the spectrum. 

Although the mechanism instigating the fracture differs, the excitation of torsional modes in GRB 211211A's precursor may be exactly analogous to that in SGRs, such as SGR 1900+14, where QPOs at frequencies of $28 \pm 2$~Hz, $53 \pm 5$~Hz, $\sim$84~Hz, and $155 \pm 6$~Hz were detected in the tail of the 1998 giant flare \citep{hur99,str05}. We explore this scenario in this short paper. We complement our study with an analysis of the X-ray afterglow of the main event (Appendix \ref{sec:afterglow}), which we find is well fitted by a magnetar wind model \citep{rowl13}. {While tangential to the main goals of the paper, this allows us to infer additional information about the event as a whole.  }

\section{Precursor flare in GRB 211211A}
\label{sec:precusors}

The precursor in GRB 211211A was observed by Fermi/GBM, Swift/BAT, and Insight-HXMT/HE \citep{xiao22}. The waiting time of the precursor relative to the main event is found to be $\sim$0.93s for Fermi, $\sim$0.88s for Swift, and $\sim$1.28s from Insight data. Overall, therefore, we consider the waiting time to be $1.08 \pm 0.20\, \mbox{s}$, which is not unusual for precursors in other merger-driven GRBs \cite[e.g. GRBs 140209A and 181126413;][]{wang20}. {With a binary-based kilonova model, \cite{rast22} find best-fit masses (see their Fig. 7) for the progenitors to be $M_{1} = 1.43 \pm 0.05 M_{\odot}$ (primary) and $M_{2} = 1.25 \pm 0.03 M_{\odot}$ (companion).} The timing delay therefore corresponds to an orbital frequency of $\Omega^{\rm pre}_{\rm orb} \approx 400 \pm 40\, \mbox{Hz}$, though it could be higher if the jet-formation timescale $\tau_{\rm jet} \gg 0$ \citep{zhang19}. 

\subsection{Resonant shattering}
\label{sec:shatter}

A neutron star, spiralling around a companion, experiences a tidal field that perturbs its crustal matter. If the frequency of the perturbing force matches the frequency of a given QNM, said mode becomes resonantly excited by efficiently depleting energy from the orbit. The amount of energy that can be transferred into the mode depends on its frequency and the degree of `tidal overlap', which defines the coupling strength to the external tidal field \cite[see, e.g.][]{alex87,kostas95}. If the overlap integral is large enough, the induced strain on the crust may exceed the elastic maximum, resulting in a failure \citep{tsang12,suvk20,kuan21b,tsang21}. {In brief, a resonant shattering is a feasible explanation for a given precursor if the following conditions are met. (i) The mode eigenfrequency should be approximately twice that of $\Omega^{\rm pre}_{\rm orb}$ since the $m=2$ harmonic of the tidal field couples most strongly to stellar perturbations \citep{alex87}. (ii) The tidal overlap integral must be large enough, such that the mode is able to reach a saturation amplitude that can break the crust. (iii) Enough energy must be able to be relieved from the crust, so as to match the precursor luminosity. These criteria are referred to in subsequent sections. The spectrum of the precursor is also important to consider, as described in the next section.}

\subsection{A magnetar in the merger?}
\label{sec:magmerg}

The precursor showed no evidence for a thermal component \citep{xiao22}. As discussed by \cite{tsang12}, this suggests that a magnetar participated in the merger. Essentially, energy is able to propagate as Alfv{\'e}n waves along open field lines if $B$ is large enough ($\gg10^{13}\, \mbox{G}$), and the spectrum will be predominantly non-thermal. \cite{xiao22} also find that the peak isotropic luminosity of the precursor is $7.4^{+0.8}_{-0.7} \times 10^{49}\, \mbox{erg s}^{-1}$, which is rather high relative to other precursors \citep{wang20}. \cite{tsang13} shows that the maximum luminosity, which is extractable from the crust of a star with radius $R$ via the magnetic field, is given by the surface integral 
\begin{equation} \label{eq:lmax}
\begin{aligned}
L_{\rm max} &= \int \left(\boldsymbol{v} \times \boldsymbol{B} \right) \times \boldsymbol{B} \cdot d \boldsymbol{A}\\
&\sim 10^{49} (v/c) (B_{\text{surf}} / 10^{14}\, \mbox{G})^2 (R / 10\, \mbox{km})^{2}\, \mbox{erg s}^{-1},
\end{aligned}
\end{equation}
where $\boldsymbol{v}$ is the velocity of the mode-related perturbation. {Demanding that $L_{\rm max} \gtrsim 7 \times 10^{49}\, \mbox{erg s}^{-1}$ implies a magnetic field of at least a few times $10^{14}\, \mbox{G}$.} The non-thermal character of the precursor and the luminosity budget \eqref{eq:lmax} thus both seem to demand magnetar-level fields for the flaring component in GRB 211211A. {As fields of this strength can non-trivially shift the QNM spectrum, as can rotation, the timing of the precursor relative to the main event (criterion i) may favour certain mode scenarios over others \citep{suvk20}.} %

\subsection{g-mode resonances}
\label{sec:gmodes}

Fluid elements in a neutron star experience buoyancy whenever away from equilibrium if there are compositional or thermal gradients, or state changes \cite[involving, e.g. superfluidity or superconductivity;][]{pass16}. These effects give rise to a rich spectrum of $g$ modes, with Brunt-V{\"a}is{\"a}l{\"a} frequencies that depend on the Fermi energies of the various particle species and the internal temperature. {In principle, $g$-mode resonances can accommodate a wide range of precursor timings. In this work,} these frequencies are calculated by assuming the adiabatic indices of the perturbed star and (normal fluid) equilibrium have a fixed ratio, which we take to be a `canonical' value appropriate for a mature star (see Appendix \ref{sec:appendix} for details). \cite{kuan21b} show that $\gtwo$ modes in this case can become resonant $\sim 1\, \mbox{s}$ prior to coalescence, exerting crust-breaking levels of strain if the stellar parameters lie in a particular range. For a wide library of equations of state (EOS), however, $\gtwo$ modes for stars with $M \approx 1.4 M_{\odot}$ tend to be `tidally-neutral', in the sense that the overlap integrals are comparatively small \cite[see Appendix A of][]{kuan22}. {It is therefore more difficult, at least within a $g$-mode resonance scenario (cf. Sec. \ref{sec:otherscenarios}), to accept a crust breaking for the primary (criterion ii violated).} 

{Assuming $B \gtrsim 10^{14}\, \mbox{G}$, the magnetic geometry (e.g. toroidal fields and multipoles) is an important ingredient for determining mode properties.} For simplicity, we adopted a dipole field, the specifics of which are given in Appendix \ref{sec:appendix}. {An additional input we required is the hydrostatic background, described by the Tolman-Oppenheimer-Volkoff equations for some EOS, with a separate model for the crust applied at low densities. We used the \cite{apr4} (APR4) EOS for the core, together with a \cite{dh01} fit for the crust. This is supplemented with the shear modulus, $\mu = \left(0.015 + 0.611 \rho_{14} + 0.599 \rho_{14}^2 - 0.349 \rho_{14}^3 + 0.273 \rho_{14}^4 \right) \times 10^{30} \text{erg cm}^{-3}$, with $\rho_{14}$ being the mass density in units of $10^{14}\, \mbox{g cm}^{-3}$, which was found by fitting the high charge-number (Z) profile calculated by \cite{tews17} (see Fig. 6 therein).}

Following \cite{kuan21b,kuan22}, we find that a $\gtwo$-mode resonance can fracture {(criterion ii)} a static companion at the right time {(criterion i;} assuming $\tau_{\rm jet} \approx 0$) for {polar} field strength $B_p=7.78_{-0.33}^{+0.41}\times 10^{14}\, \mbox{G}$ {(criterion iii)}, where the uncertainties account for the $\sim 0.2\, \mbox{s}$ timing window \citep{xiao22}; readers can refer to Fig. \ref{fig:g2shatter} for more details. The formation of a binary with such a component is unlikely \citep{pop06}, though it cannot be ruled out {a priori}, {and it is consistent with the luminoisty and non-thermality considerations detailed in Sec. \ref{sec:magmerg}.}

\subsection{Other resonance scenarios}
\label{sec:otherscenarios}
{Our main goal in this work is to demonstrate that a tidal resonance, followed by aftershocks, can explain the observations leading up to GRB 211211A. The former can be accomplished with a $\gtwo$ mode for a normal fluid star, of mass equal to that of the merger companion ($M = 1.25 M_{\odot}$) with a dipole field of strength $B_{p} \sim 8 \times 10^{14}\, \mbox{G}$. This is not, however, the only viable scenario. For example, \cite{tsang21} show that, depending on the nuclear symmetry energy and other properties of the star, the frequency of crustal $i$ modes span a large range between $50 - 450$ ($2 \pi$ Hz), which could easily match $2 \Omega^{\rm pre}_{\rm orb}$. In non $g$-mode interpretations, one also cannot exclude the possibility of the primary being the flarer. Detailed, {ab initio} simulations of excitation and fracture are needed to tell resonance scenarios apart since the basic criteria outlined in Sec. \ref{sec:shatter} can be met with several models. We defer such considerations to future work (though see also Secs. \ref{sec:resonant} and \ref{sec:disprove}).
}

\section{Seismic excitations}
\label{sec:resonant}
\cite{xiao22} identified, with $> 5 \sigma$ significance, QPOs with frequency $22^{+3}_{-2}$Hz in the GRB 211211A precursor. Secondary Fourier peaks are also visible at $51 \pm 2\, \mbox{Hz}$ in the Fermi spectrum \cite[see extended data Figures 7 and 8 in][]{xiao22}. If QNMs do indeed soak up enough orbital energy, such that the resulting crustal strain exceeds the threshold of the lattice structure in the crust $\sim 1\, \mbox{s}$ prior to merger, aftershocks may excite torsional modes. Such a scenario is qualitatively similar to that put forth by \cite{duncan98} for SGR flares, where similar frequencies have been observed \citep{hur99,is05}.

{Magnetar QPOs are now understood to be more complicated, as crust and core oscillations are strongly coupled by the magnetic field. Similar considerations likely apply to the flaring component in GRB 211211A, as argued in Sec. \ref{sec:magmerg}. The total spectrum for axisymmetric stars with poloidal fields contains multiple (Alfv{\'e}n) continua, separated by `gaps' where discrete crustal and global modes reside \cite[e.g.][]{vh11}. If the field is tangled \citep{link16}, polar-axial couplings are allowed \citep{col12b}, or if axisymmetry is broken \citep{sot11}, the continua dissolve into discrete modes. Either way, it is this richness that allows for the multiple, close QPO frequencies observed in the giant flare from SGR 1806--20 \cite[$\sim$16, 18, 22, 26, ..., Hz;][]{ham11}, for instance, to be modelled through QNMs. \cite{col12} found that $\sim22$Hz oscillations are most easily explained with a crustal mode, residing on the edge of the Alfv{\'e}n spectrum (see Fig. 1 therein). As this frequency coincidentally matches that reported by \cite{xiao22} for GRB 211211A, we focus on torsional modes here. It should be stressed though that the detection of only one or two frequencies makes it difficult to ascertain the mode character(s) (cf. Sec. \ref{sec:disprove}). For example, \cite{gab16} find that coherent, global magnetoelastic modes in a superfluid star with $B \sim 5 \times 10^{14}\, \mbox{G}$ also share a frequency of $22$Hz. Superfluidity, however, increases $g$-mode frequencies \citep{pass16}, and thus a $\gtwo$ mode is no longer a viable candidate for precursor ignition in this case. Though a magnetar likely contributed to GRB 211211A, resonant shattering events do not necessitate magnetar involvement in general, especially for thermal or dim precursors (see Sec. \ref{sec:magmerg}). Pure crustal modes would be the most natural explanation in such cases.} 

The strain tensor associated with the Lagrangian displacement of the resonant mode(s) can be used to deduce the fracture geometry, given a chemical model of the crust \citep{suvk20,kuan21b}. The resulting spatial profile can be projected into spheroidal harmonics, which serve as initial data for a {QNM} problem. In this way, predictions for seismic {or magnetoelastic} modes can be matched to fracture geometries.

\subsection{Quasi-periodic oscillations from aftershocks}
\label{sec:qposec}

%A first-principles analysis, as described in the previous section, lies beyond the scope of this paper. 
{As a first step to solving the full problem described above,} we made some quantitative predictions by computing the properties for free (i.e. without fracture seeding) torsional modes, assuming they are excited immediately after a tidal resonance. If the yielding is caused by a $\gtwo$ mode (see Sec. \ref{sec:gmodes}), the cracking area\footnote{The crack geometry also depends on the failure type (e.g. von Mises or Tresca) and the (likely anisotropic) elastic maximum the crust can absorb, neither of which are certain for neutron star matter; see \cite{baiko17} for a discussion.} is confined near the equator, with colatitudes $0.25 \lesssim \theta/\pi \lesssim 0.75$ \cite[see Fig.~3 in][]{kuan21b}. The angular distribution is similar to that of $\ell =m = 2$ harmonics. Modes with azimuth numbers $m\ne2$ are largely decoupled from the perturbation due to mode orthogonality. Similarly, odd-order harmonics with $\ell = 3, 5, \ldots$ are dormant because the $\gtwo$ mode seeding the fracture is reflection symmetric about the equator. For GRB 211211A then, the modes of most interest are $\tor$ and $\torf$ (both with $m=2$; {though cf. the caveats noted in Sec. \ref{sec:resonant}).}

\subsection{Torsional mode spectrum}

To calculate the torsional modes, we closely followed the formalism described by \cite{sot07}; in particular, their equations (69) and (70) together with boundary conditions (71) and (72) completely describe the eigenproblem, including magnetic corrections, for torsional modes in the Cowling approximation. {We solved these equations in this work using the EOS, shear modulus, and magnetic field specifications given in Sec. \ref{sec:gmodes}.}

We find $\tor^{(0)} = 24.22\, \mbox{Hz}$ and $\torf^{(0)} = 51.35\, \mbox{Hz}$ for a static, unmagnetised star with $M = 1.25 M_{\odot}$. These already agree well with the Fourier peaks observed by \cite{xiao22}, though they shift slightly if rotational and magnetic corrections are applied. For $m=2$, we find a hybrid fitting to the frequency shift in the form
\begin{equation} \label{eq:cor}
\frac{{}_{\ell}t_{n}} {{}_{\ell}t^{(0)}_{n}}  = \sqrt{1+\left( \frac{B_{p}}{{{}_{\ell}\tilde{B}^{}_{n}}} \right)^{2}}- \frac{2 \nu(\ell^2+\ell-1)}{{}_{\ell}t^{(0)}_{n}  \ell(\ell+1)} + \mathcal{O}\left[ \left( \frac{B_{p}}{{{}_{\ell}\tilde{B}^{}_{n}}}\right)^2, \left(\frac{\nu}{{}_{\ell}t^{(0)}_{n}} \right)^2 \right] ,
\end{equation}
for fitting coefficients ${{}_{2}\tilde{B}^{}_{0}} = 2.78 \times 10^{15}\, \mbox{G}$ and ${{}_{4}\tilde{B}^{}_{0}} = 2.65 \times 10^{15}\, \mbox{G}$, where ${}_{\ell}t_n$ is the inertial-frame frequency when $\nu > 0$. Equation \eqref{eq:cor} is similar to Eq.~(79) of \cite{sot07} when $\nu = 0$, and Eq.~(26) of \cite{vav07} when $B_{p} = 0$.

\subsection{QPOs in the precursor of GRB 211211A}

A variety of $\ell =2 $ and $\ell =4$ torsional mode frequencies are collated in Table \ref{tab:mode}.  Although the $\nu = B_{p} = 0$ mode frequencies match those of the QPOs in GRB 211211A, a $\gtwo$-mode resonance is unable to explain the precursor in this case, {as the spectrum would be thermal and criteria (i) and (iii) would be violated (see Sec. \ref{sec:precusors}).} In contrast, we find that there are many possible combinations of modest spins, $0 \lesssim \nu/\mbox{Hz} \lesssim 3$, and polar field strengths, $7 \lesssim B_{p}/10^{14}\mbox{G} \lesssim 8$, that {can} allow for a $\gtwo$-resonant shattering {to meet all necessary criteria}, and excite torsional modes with the observed frequencies. Three such combinations are shown in the last three rows of Tab. \ref{tab:mode}. 

\begin{table}
        \centering
        \caption{Frequencies (third column) of nodeless torsional modes in the companion ($M_{2}=1.25M_{\odot}$), for a range of spins $\nu$ (first column) and polar field strengths $B_{p}$ (second column) via expression \eqref{eq:cor}.}
        \label{tab:mode}
        \begin{tabular}{ccc} 
                \hline 
                $\nu$ (Hz) & $B_{p}$ ($10^{13}\, \mbox{G}$) & Mode Freq. ($\ell =2 ,  \ell = 4$) (Hz) \\
                \hline 
            0 & 0 & $\boldsymbol{24.22}$  \qquad$\boldsymbol{51.35}$  \\                
            2 & 0 & $\boldsymbol{20.89}$\qquad  47.55  \\               
            4 & 0 & 17.55\qquad 43.75  \\       
            \hline
            0 & 1 & $\boldsymbol{24.22}$\qquad  $\boldsymbol{51.35}$  \\                
            0 & 10 & $\boldsymbol{24.24}$\qquad  $\boldsymbol{51.39}$  \\               
            0 & 100 & 25.74  \qquad 54.88 \\    
            \hline      
            $\boldsymbol{0.5}$ & $\boldsymbol{80}$ & $\boldsymbol{24.37}$  \qquad$\boldsymbol{52.69}$ \\
            $\boldsymbol{1}$ & $\boldsymbol{76}$ & $\boldsymbol{23.44}$  \qquad$\boldsymbol{51.52}$  \\
            $\boldsymbol{2}$ & $\boldsymbol{72}$ & $\boldsymbol{21.69}$  \qquad$\boldsymbol{49.41}$  \\
            \hline
        \end{tabular}
        \tablefoot{The core EOS is APR4 \protect\citep{apr4}, while the \protect\cite{dh01} fit governs the crust, which occupies the outermost $0.973$ km of the star; such a model is able to explain the QPOs observed in the giant flare tails of SGR 1806--20 and SGR 1900+14 \citep{col12}. Bold numbers match the QPOs in GRB 211211A's precursor, and/or the spin and magnetic field strengths predicted if a $\gtwo$-mode resonance breaks the crust $\sim 1.08 \pm 0.20\, \mbox{s}$ prior to coalescence (see Sec. \ref{sec:shatter}).}
\end{table}

%\subsection{}

\section{Discussion}
\label{sec:discussion}

If resonant shatterings are responsible for (at least some) precursor flares in merger-driven GRBs, seismic aftershocks may emanate from the rupture site shortly thereafter. We show that the $\sim 22$ and $\sim 51\, \mbox{Hz}$ Fourier peaks observed in the precursor of GRB 211211A match those of $\ell =2$ and $\ell =4$ torsional modes for a variety of spin and magnetic field strength combinations (Tab. \ref{tab:mode}). These modes are expected based on the fracture geometry of $\gtwo$-mode resonances, as quantified in Sec. \ref{sec:qposec}. Even if a different mode is responsible for the resonant shattering \cite[e.g. $i$ mode;][]{tsang12}, the same qualitative picture applies, {though the specific fracture patterns and timings may differ (see Sec. \ref{sec:otherscenarios}).} Overall, the observation of precursor QPOs may allow for detailed inferences regarding the flaring star's properties.

\subsection{Avenues to disprove the scenario}
\label{sec:disprove}
{An important question to ask is how might one falsify the resonant aftershock picture. For the resonance itself, because $g$ and $i$ modes are so sensitive to the stellar composition, precursor timings from $\lesssim1\,\mbox{s}$ to even $\gtrsim10\,\mbox{s}$ prior to the main event can be met with some choice of mass, EOS, spin, and magnetic field strength \citep{kuan21a,tsang21}. It is only when several (or all) of these aspects are pinned down -- the first and last of which are for GRB 211211A -- can models be definitively ruled out. For example, because of the tidal neutrality phenomenon for $\gtwo$ modes discussed in Sec. \ref{sec:gmodes}, if it were proven that the {primary} released the flare, the relevant overlap integral would not be large enough to fracture the crust (criterion ii violated), and a $\gtwo$-mode scenario could be excluded.}

{Torsional mode frequencies do not match the QPOs if the star is spinning at even a few Hz (see the third row of Tab. \ref{tab:mode}). Galactic magnetars spin slowly, implying that torsional modes take on a relatively narrow range of properties in such stars, but high spins cannot be ruled out {a priori} for the components in a merger. If gravitational waves are observed coincidentally with a GRB and its precursor, it may be possible to measure the stellar spin(s) \citep{lasky18}, thereby offering an avenue to exclude torsional modes. Excluding all QNM models (e.g. global magnetoelastic modes) is difficult, as again a wide range of QPO frequencies can be accommodated for various magnetic geometries \citep{col12}, or if superfluidity is considered \citep{gab16}. If only high-frequency ($\gtrsim 10^{2}\, \mbox{Hz}$) QPOs were observed in a precursor, however, this would exclude an aftershock picture since low-order modes are easiest to excite, and their absence would imply a non-fracture mechanism unless the star is ultra-magnetised \cite[$B_{p} \gtrsim 4 \times 10^{16}\, \mbox{G}$; see Fig. 4 of][]{sot07}.} Such a star would be very difficult to explain in a neutron star merger event as field decay would occur on $\ll$Myr timescales, long before coalescence.

\subsection{Future prospects}
{Given that giant flare tails are visible for $\gtrsim 10^{2}$s \citep{ws06}, while GRB precursor emissions last less than a second \cite[$\sim 0.1$s for GRB 211211A;][]{xiao22}, it is not surprising that fewer Fourier peaks are observed in the latter. With improved instruments and more observations of precursor QPOs, additional modes may be detected in the future. In general, one would expect many QNMs, provided they can activate within $\lesssim 0.1$s \cite[cf. Fig 10 in][]{stroh06}, to be excited in a rupture event beyond just $\tor$ and $\torf$ in a magnetised star. With each peak comes a constraint on the multi-dimensional parameter space relating the mass, spin, magnetic field strength, shear modulus, and so on, to the QNM spectrum \citep{col12b}. For GRB 211211A, we have only one confidently measured frequency, and thus many possibilities are defensible, even within the restricted context of torsional modes (Tab. \ref{tab:mode}). If many QPOs spanning $\sim15$--$1500\, \mbox{Hz}$ were detected in a precursor, such as in the tail SGR 1806--20's giant flare, this would tightly constrain the active member of the coalescing binary.}

{The above can be combined with constraints coming from gravitational waves, the kilonova, and afterglows \citep{gomp22}. We show in Appendix \ref{sec:afterglow} that the X-ray afterglow for GRB 211211A indicates that the remnant is a rapidly rotating, (meta-)stable magnetar with $B_{p} \gtrsim 10^{14}\, \mbox{G}$. This and other (e.g. tidal deformability) properties can, at least in principle, be connected to those of the pre-merging objects \citep{radice18}. Such considerations are relevant for population studies and evolutionary modelling of neutron stars \citep{pop06}.}

\begin{acknowledgements}
The research leading to these results has received funding from the
European Union's Horizon 2020 Programme under the AHEAD2020 project (grant n. 871158). This work made use of data supplied by the UK Swift Science Data Centre at the University of Leicester. KK gratefully acknowledges financial support by DFG research Grant No. 413873357. We are grateful to the anonymous referee for providing helpful and specific feedback, which improved the quality of this work.
\end{acknowledgements}

\bibliographystyle{aa}

\appendix

\section{Afterglow}
\label{sec:afterglow}

The components in GRB 211211A are relatively light ($M_{1} + M_{2} \approx 2.68 \pm 0.08 M_{\odot}$), suggesting the possibility of a neutron star merger remnant, rather than a black hole. From a three-component kilonova model, \cite{rast22} estimate that $0.04 \pm 0.02 M_{\odot}$ worth of $r$-process material was dynamically ejected via tidal torques and shock heating during a merger. After coalescence, additional mass may be lost as material in the temporary accretion disc surrounding the merger site unbinds through viscous heating. \cite{fuji18,fuji20} show that, for a long-lived remnant, the post-merger ejecta mass can reach $\approx 30-40\%$ of the total disc mass, amounting to $\gtrsim 0.1 M_{\odot}$. These considerations suggest a remnant mass of $M_{\text{rem}} \approx 2.5 \pm 0.1 M_{\odot}$. A star at the lower end could stave off collapse for a number of spindown timescales if born with high angular momentum \cite[e.g.][]{ravi14}, or even indefinitely, depending on the nuclear EOS.

{GRB 211211A displayed an X-ray afterglow, part of which could be supplied by a magnetar wind \citep{rowl13,cio18}.
The photon index for X-ray emissions and source redshift were determined as $\Gamma_{\gamma} = 2.08$ and $z = 0.0763,$ respectively, the latter implying a distance of $d \approx 350\, \mbox{Mpc}$ \citep{ev14,gomp22}. Assuming uncollimated emissions, we translated the observed flux into an isotropic luminosity, namely $L_{\rm X} = 4 \pi d^2 K(z) F$ for $K(z) \approx (1+ z)^{\Gamma_{\gamma}-2}$ \citep{bloom02}. The electromagnetic spindown luminosity of an oblique rotator braking via magnetic dipole torques is given by
\begin{equation} \label{eq:spindown}
L_{\rm sd} \approx -I_{\star} \Omega \frac {d \Omega} { dt} = \eta \frac {(B^{\star}_{p})^2 R_{\star}^{6} \Omega^4} {6 c^3} \lambda(\alpha),
\end{equation}
where $B^{\star}_{p}$ denotes the polar field strength, $R_{\star}$ the radius, $\Omega = 2 \pi \nu_{\star}$ the angular velocity, $c$ the speed of light, $I_{\star}$ the moment of inertia, and $\lambda$ is a magnetospheric factor that depends on the inclination angle, $\alpha$. Assuming the source is surrounded by a vacuum and that $\alpha \approx \pi/2$, we have $\lambda \approx 1$; similar values are expected if $\alpha \approx 0$, but the star is surrounded by plasma \cite[see, e.g.][]{rowl13,cio18}. The dimensionless prefactor, $\eta$, is an efficiency term, which controls how easily spindown power is converted into X-rays. Though theoretically uncertain, \cite{xiao19} analysed several GRB lightcurves to find that $10^{-7} \lesssim \eta \lesssim 10^{-1}$, depending on the saturated Lorentz factor of the emissions (see their Fig. 2). We consider an intermediate value here, $\eta \approx 10^{-4}$. Fitting expression \eqref{eq:spindown} to $L_{\rm X}$, we find
\begin{equation} \label{eq:bfield}
B^{\star}_{p} = 4.6^{+1.3}_{-1.0} \times 10^{14} \left(\frac {I_{\star}} {3 \times 10^{45} \text{g\,cm}^{2}} \right) \left( \frac {\eta} {10^{-4}} \right)^{1/2} \left( \frac {11 \text{km}} {R_{\star}} \right)^{3} \mbox{G}
\end{equation} 
and 
\begin{equation} \label{eq:spin}
\nu_{\star} = 1.59^{+0.26}_{-0.27} \left(\frac {3 \times 10^{45} \text{g\,cm}^{2}}  {I_{\star}} \right)^{1/2} \left( \frac {10^{-4}} {\eta} \right)^{1/2} \mbox{kHz},
\end{equation}
for the {remnant's} polar field strength and spin frequency, respectively, at a 95$\%$ confidence level ($\chi^2/{\rm dof} = 53.9/49$, weighting for uncertainties).}

\begin{figure}
\includegraphics[width=\columnwidth]{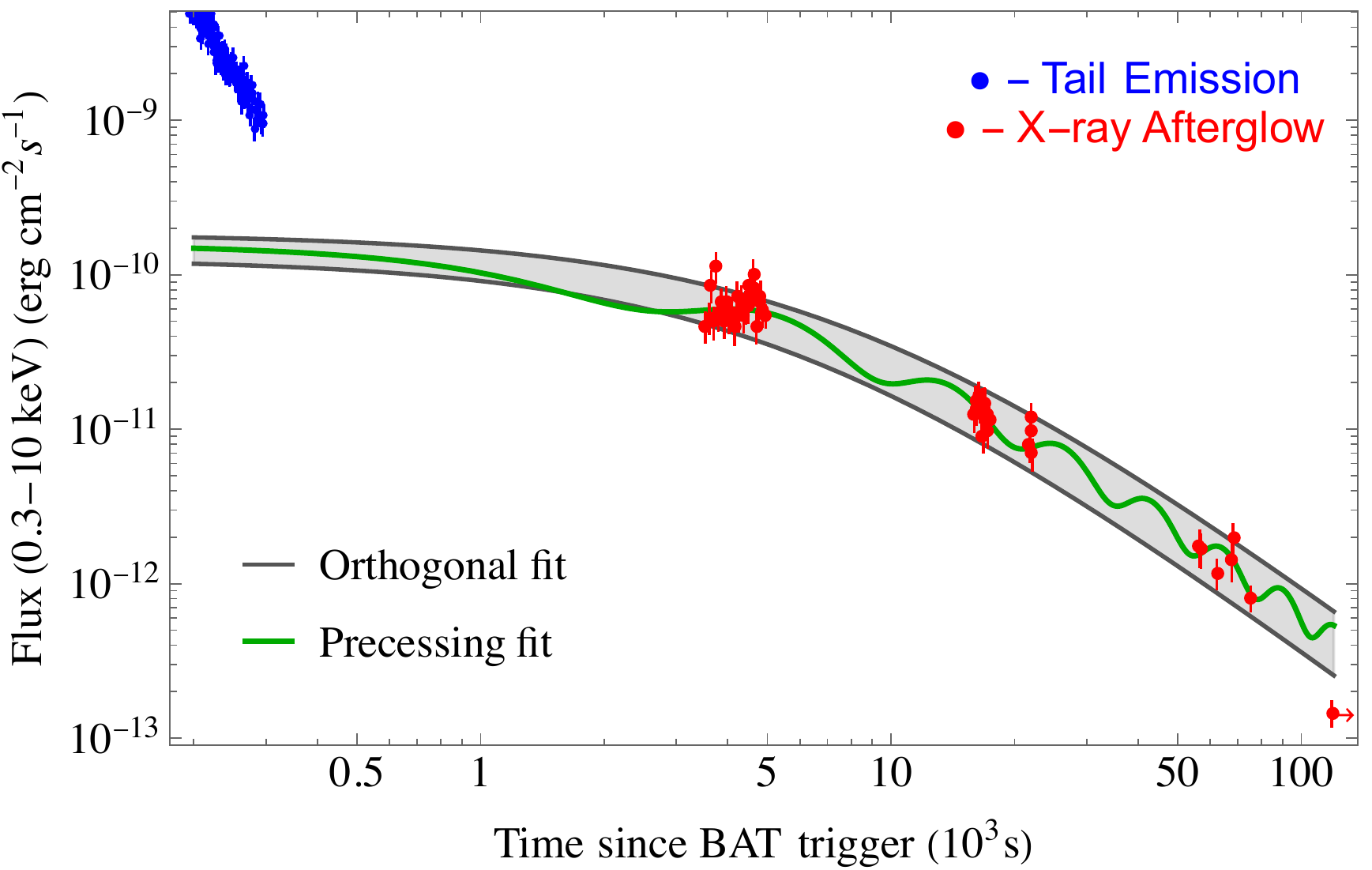}
\caption{Afterglow light-curve fittings for GRB 211211A. Blue data are attributable to the decaying tail of the prompt emission \protect\citep{ev14}, while red data may be contributed by a newborn magnetar. The grey band shows the $95\%$ confidence interval for an orthogonal-rotator fit, while the green curve shows a best-fit precessing model. }
\label{fig:afterglow}
\end{figure}

{From numerical merger simulations with the APR4 EOS, \cite{radice18} find that the spin period for a newborn remnant is well fitted by $P_{\star} \approx \left[ -0.21 \left( M_{\rm rem}/M_{\odot} - 2.5 \right) + 0.67 \right]\, \mbox{ms}$, which agrees with expression \eqref{eq:spin}. However, the APR4 EOS, for example, can only sustain maximum masses up to $\approx 2.21 M_{\odot}$ if the star is static. This limit increases by $\gtrsim 20\%$ if the star is rotating near the break-up limit \cite[e.g.][]{suvg22}. Assuming a remnant of mass $M_{\rm rem} \approx 2.4 M_{\odot}$, this implies that the star would collapse after $\sim 10\, \mbox{ks}$ for magnetic field strength \eqref{eq:bfield} and spin \eqref{eq:spin} -- when centrifugal support is depleted by the magnetar wind \cite[see Eq.~(2) in][]{ravi14} -- unless $B_{p}$ or $\eta$ are on their respective low ends, spindown is delayed (through fallback-induced spinup, for instance), or the excess mass is magnetically supported by a strong core field \citep{suvg22}. This is potentially at odds with the fact that the flux persisted for $\gtrsim 50\, \mbox{ks}$ post-trigger (see Figure \ref{fig:afterglow}), as black hole formation is expected to truncate the signal. \cite{ravi14} show that this could, in principle, allow us to rule out EOS candidates that have low (static) maximum masses. We emphasise, however, that the remnant mass estimate is itself model dependent, and thus strong conclusions cannot be drawn from this event alone \citep{fuji18,fuji20,rast22}.} 

In addition to an orthogonal-rotator fit, we consider the possibility of a precessing remnant for completeness. Although the model specifics are complicated in this case \cite[see][for details]{suvk21a,suvk21}, such a fit predicts similar parameters to the orthogonal case; we find $B_{p} \approx 4.4 \times 10^{14}\, \mbox{G}$, $\nu \approx 1.55\, \mbox{kHz,}$ and a quadrupolar ellipticity of $\epsilon \approx 1.6 \times 10^{-7}$ for the same efficiency and moment of inertia. %($\chi^2/\{\rm dof} = 45.3/46$).%, $I_{\star}$.   
Orthogonal (black) and precessing (green) fits to the afterglow light curve of GRB 211211A are shown in Fig. \ref{fig:afterglow}. Overall, a magnetar wind model fits the data well, as they exhibit a plateau followed by a power-law fall-off with slope $-2$, as predicted by dipole braking. 

\section{Magnetic field and spin inferences from precursor timing}
\label{sec:appendix}
As discussed in Secs. \ref{sec:precusors} and \ref{sec:resonant}, mode frequencies are shifted by Coriolis and Lorentz forces. Quantifying these effects requires a particular model; the one used in the main text is detailed here. The relative shifts also depend on the mode eigenfunction. For torsional modes, the shifts are given by expression \eqref{eq:cor} in the main text, while for $g$ modes the patterns are different. In either case, specifying the nuclear EOS -- assumed APR4 in this work because \cite{col12} found that torsional modes in a $M \approx 1.4 M_{\odot}$ star with this EOS match the QPO frequencies in the giant flare tails of SGR 1806--20 and SGR 1900+14 \citep{is05} -- uniquely determines the former if the star is uniformly rotating. Relative to a static-star mode, we find that the leading-order, rotational corrections to $m=2$ modes, ${}_{\ell}\omega_{n}$, in geometrised units with $G = c = 1$, read [see Eqs.~(17) and (18) in \cite{kuan21b}]
        \begin{align}\label{eq:modrot}
                {}_{\ell}\delta\omega^{\rm Cor}_{n}= -4 \pi \nu (1-C_{n \ell}),
        \end{align}
        with
        \begin{align}
                C_{n \ell}= \frac{1}{M R^2}\int dr (p + \rho) e^{\Phi+\lambda}r^{2\ell-2}
                 (\xi_{r}\overline{\xi}_{h} + \overline{\xi}_{r}\xi_{h} + \xi_{h}\overline{\xi}_{h}),
        \end{align}
for mode eigenfunction $\boldsymbol{\xi}$, decomposed into radial ($\xi^r$) and poloidal ($\xi^h$) harmonics, stellar pressure $p$ and density $\rho$, related to the metric functions $\Phi$ and $\lambda$, defining the (logarithms of the) $g_{tt}$ and $g_{rr}$ components, respectively, via the Tolman-Oppenheimer-Volkoff relations. In the Newtonian limit, the above agrees with the classic Unno formula.

Although the presence of multipoles or a toroidal field can lead to sizeable adjustments in the Lorentz-related frequency shift \cite[see, e.g.][]{col12,suvk20,kuan21a}, a purely dipole configuration for the magnetic field is considered here for simplicity. Following \cite{kuan21a}, it can be shown that relations of the Faraday tensor imply that a general, stationary, and axisymmetric magnetic field has contravariant components
\begin{equation} \label{eq:magf}
B^{\mu} = \frac{B_{p}}{2} \left( 0, \frac {e^{-\lambda}} {r^2 \sin\theta} \frac{\partial\psi}{\partial\theta}, 
- \frac {e^{-\lambda}} {r^2 \sin\theta}  \frac{\partial\psi}{\partial r}, 
- \frac {\zeta(\psi) \psi e^{-\Phi}} {r^2 \sin^2\theta}  \right)\end{equation}
in Schwarzschild-like coordinates, where $\zeta$ is a toroidal function, which we set to zero. For a dipole field, the streamfunction $\psi$ can be expressed as
\begin{equation}
\psi(r,\theta) = f(r) \sin^2\theta\end{equation}
for some radial function $f$. We picked\footnote{To avoid confusion, we note that the function $f(r)$ is denoted by $a_1(r)$ in \cite{sot07}, which affects the spectrum of torsional modes via the associated terms in Eqs.~(69) and (70) therein.} $f(r) = a_{1} r^2 + a_{2} r^4 + a_{3} r^6$, where the $a_{i}$ are constrained by assuming (I) that the $\boldsymbol{B}$ field smoothly extends to a force-free dipole outside of the star (where the geometry is Schwarzschild), and (II) that there are no surface currents. Enforcing these conditions leads to
\begin{subequations}
\begin{align}
a_{1} =& - \frac {3 R^3} {8 M^3} \Bigg[ \log \left( 1 - \frac{ 2 M} {R} \right) \nonumber\\
&+ \frac {M \left( 24 M^3 - 9 M^2 R - 6 M R^2 + 2 R^3 \right)} {R^2 \left( R - 2 M \right)^2} \Bigg], \label{eq:a1}\\
a_{2} =& \frac {3 \left( 12 M - 7 R \right)} {4 R \left( R - 2 M \right)^{2}}, \label{eq:a2}
\end{align}
and
\begin{align}
a_{3} = \frac {3 \left( 5 R - 8 M \right)} {8 R^3 \left( R - 2 M \right)^{2}}. \label{eq:a3}
\end{align}
\end{subequations}
Given this field, the Lorentz force $F^{\rm Lor}_{\mu}$ can be defined, which leads to a frequency shift in the unmagnetised value, in other words
\begin{align}\label{eq:freqpert}
{}_{\ell}\delta\omega^{\rm Lor}_{n} =  \frac{1}{2} 
\frac{{\int_{\text{vol}} F^{\rm Lor}_{\mu}\bar{\xi^{\mu}} \sqrt{-g}d^{3}x}}
{{}_{\ell}\omega_{n}{\int_{\text{vol}}(\rho+p)e^{-2\Phi}\xi^{\mu}\bar{\xi_{\mu}}\sqrt{-g}d^{3}x}}.
\end{align}

%%%%%%%%%%%%%%%%%
\begin{figure}
\includegraphics[width=\columnwidth]{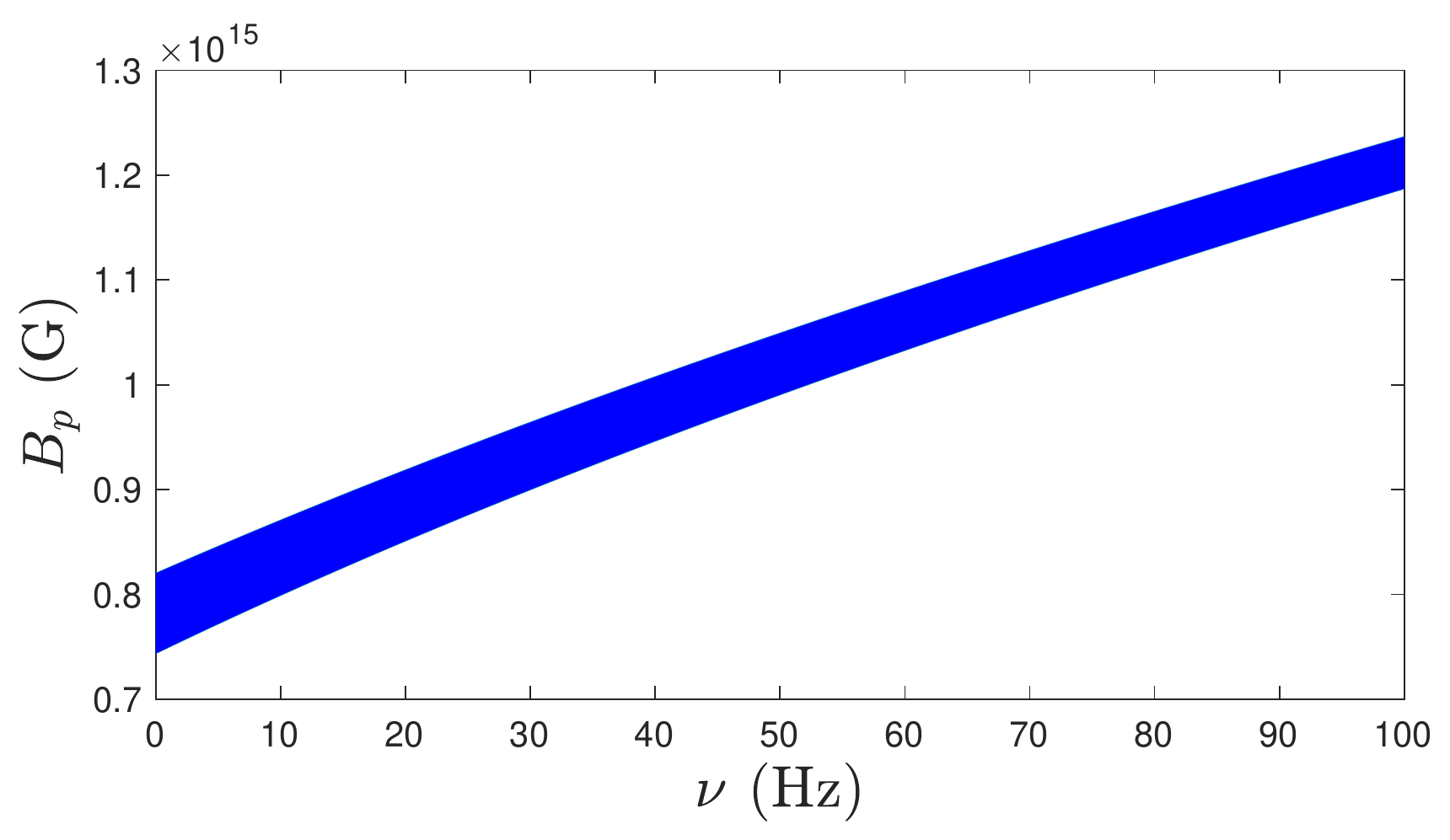}
\caption{Polar magnetic field strengths $B_p$, as a function of spin $\nu$, such that the resonance of a $\gtwo$ mode is able to instigate a crust yielding at the observed preceding time, $T_{\rm wait} = 1.08 \pm 0.20\, \mbox{s}$, assuming a negligible jet-formation time, $\tau_{\rm jet} = 0$, in GRB 211211A. The shaded region reflects the uncertainty in the waiting time.}
\label{fig:g2shatter}
\end{figure}
%%%%%%%%%%%%%%%%%

The matching of the $\gtwo$ mode to the precursor time implies a relationship between $B_{p}$ and $\nu$ for a given mass and stratification, that is we solved $2\Omega^{\rm pre}_{\rm orb} = {}_{2} \omega_{2} + {}_{2}\delta\omega^{\rm Cor}_{2} + {}_{2}\delta\omega^{\rm Lor}_{2} = 2 \pi \gtwo$ in the inertial frame, implicity assuming spin-orbit alignment \citep{alex87,kostas95}. We assumed $M = 1.25 M_{\odot}$, as identified by \cite{rast22}, for the companion in GRB 211211A (see Sec. \ref{sec:shatter}). The stratification is encoded by a phenomenological parameter that incorporates the particle Fermi energies and internal temperature, called $\delta$ in \cite{kuan21b,kuan22}. In general, the equilibrium  adiabatic index reads
\begin{equation}
\gamma(r) = \frac {\rho + p} {p} \frac {\partial p} {\partial \rho},
\end{equation}
which we relate to the index of the perturbed star, $\Gamma$, through
\begin{equation} \label{eq:adpert}
\Gamma(r) = \gamma(1 + \delta).
\end{equation}
The frequencies of thermal $g$ modes are however much more sensitive to the surface temperature \cite[see Sec.~2.1 of][for quantifications]{kuan22}. We assumed $\delta$ is a constant that is set by the value in the crust; a `canonical' value, appropriate for a mature neutron star, was adopted, in other words $\delta = 0.005$ \cite[justified in][]{kuan22}. This eventually yields values for ${}_{2} \omega_{2}$, and the $\gtwo$-mode frequency can now be written as a function of $B_{p}$ and $\nu$. 

Figure \ref{fig:g2shatter} shows simultaneous estimates on $B_p$ and $\nu$, based on the above procedure, with the shaded area representing the uncertainty (i.e. $T_{\text{wait}} = 1.08 \pm 0.20\, \mbox{s}$). For all spin frequencies, we find that magnetar-level fields of $B_{p} \gtrsim 7 \times 10^{14}\, \mbox{G}$ are required. Fields of this strength are consistent with the precursor energetics and non-thermality, discussed in Sec. \ref{sec:discussion}. Though not shown, non-zero values for $\tau_{\rm jet}$ imply even higher field strengths.

We emphasise that, although Fig. \ref{fig:g2shatter} pertains to $\gtwo$ modes, similar graphs may be drawn for other resonant shattering scenarios \cite[involving, for example, $i$ or $s$ modes;][]{tsang12,tsang21}. Furthermore, $g$ modes, with the simple one-parameter characterisation \eqref{eq:adpert}, are only considered here for concreteness.

\end{document}